\documentclass[12pt,a4paper]{article}
\makeatletter
\def\eqnarray{%
\stepcounter{equation}%
\let\@currentlabel=\theequation
\global\@eqnswtrue
\global\@eqcnt\z@
\tabskip\@centering
\let\\=\@eqncr
$$\halign to \displaywidth\bgroup\@eqnsel\hskip\@centering
$\displaystyle\tabskip\z@{##}$&\global\@eqcnt\@ne
\hfil$\displaystyle{{}##{}}$\hfil
&\global\@eqcnt\tw@$\displaystyle\tabskip\z@{##}$\hfil
\tabskip\@centering&\llap{##}\tabskip\z@\cr}
\makeatother
\newcommand{\ket}[1]{{\vert{#1}\rangle}}
\newcommand{\bra}[1]{{\langle{#1}\vert}}
\newcommand{\braket}[2]{{\langle{#1}\vert{#2}\rangle}}
\newcommand{\bellbraket}[2]{{\langle \langle{#1}\vert \vert{#2}\rangle 
           \rangle}}
\newcommand{\bell}[1]{{\vert \vert{#1}\rangle \rangle}}

\newcommand{\fukuso}{{\mathbf C}}

\newcommand{\futon}{{\bf N}}

\newcommand{\zetta}[1]{{\vert{#1}\vert}}

\begin{document}

\title{\sl A Relation between Coherent States and 
Generalized Bell States}
\author{
  Kazuyuki FUJII
  \thanks{E-mail address : fujii@math.yokohama-cu.ac.jp}\  
  \thanks{Home-page : http://fujii.sci.yokohama-cu.ac.jp}\\
  Department of Mathematical Sciences\\
  Yokohama City University\\
  Yokohama 236-0027\\ 
  JAPAN
  }
\date{}
\maketitle\thispagestyle{empty}
%
%
%
%
\begin{abstract}
  In the first half we show an interesting relation between coherent 
  states and the Bell states in the case of spin 1/2, which was 
  suggested by Fivel. 

  In the latter half we treat generalized coherent states and try 
  to generalize this relation to get several generalized Bell states.

  Our method is based on a geometry and our task may give a hint to 
  open a deep relation between a coherence and an entanglement. 
\end{abstract}

\newpage

%
%
%
%

\section{Introduction}
The recent progress of quantum information theory including quantum 
computer, quantum cryptgraphy and quantum teleportation is marvelous enough. 
The coherence and entanglement play an essential role in quantum information 
theory. See the papers in \cite{LPS} or \cite{AH}.

In \cite{JB} Bell considered the so--called Bell states to test the EPR 
problem (``paradox'') and proposed the famous inequality, see \cite{APe} or 
\cite{AH}.
The Bell states are typical examples of the entanglement. Interestingly 
enough they have been used in the field of quantum teleportation.  
They are in the case of spin $1/2$. Of course we can consider states with 
general spin $j$. We call them generalized Bell states. 

On the other hand coherent states are fundamental tools in quantum optics 
and they are of course entangled. See \cite{MW}.  
Coherent states (generalized coherent states) are related with 
unitary representations of compact or non--compact Lie groups such as 
$U(n)$ or $U(n-1,1)$, see \cite{AP}. 

What is a relation between coherent states and Bell states or generalized 
Bell states ?  We would like to construct a mathematical theory 
between them. 
In \cite{DF} Fivel defined the generalized Bell states as the integral of 
tensor product of generalized coherent state and its ``twisted'' one. 
We redefine Fivel's one to be more calculable and perform several 
integrals. Then we recover the Bell states and, moreover, get  
Bell states with general spin and more.
In a certain sense the states of Fivel are overcomplete expression of 
Bell states or generalized Bell states. 

By the way we are now developping Holonomic Quantum Computation, 
\cite{ZR}--\cite{KF4}. 
One of our aim of this study is to apply the idea of generalized Bell states 
to it. But we have a trouble. 
The Fivel's states are not defined for coherent states based on non--compact 
Lie group such as $U(n-1,1)$. 
This point is unsatisfactory to us. Therefore we need to extend our method 
more widely.

\vspace{1cm}
\section{Review on General Theory}
We make a review of \cite{DF} within our necessity. Let $G$ be a compact 
linear Lie group (for example $G = U(n)$) and consider a coherent 
representation of $G$ whose parameter space is a compact complex manifold 
$S = G/H$, where $H$ is a subgroup of $G$. For example $G = U(n)$ and 
$H = U(k) \times U(n-k)$, then $S$ = $U(n)/ U(k) \times U(n-k) \cong 
G_{k}({\fukuso}^{n})$, a complex Grassmann manifold. See in detail \cite{AP} 
or \cite{FKS}. Let $Z$ be a local coordinate and $\ket{Z}$ a generalized 
coherent state in some representation space $V$ ($\cong {\fukuso}^{K}$ for 
some big $K \in \futon$). Then we have from the definition 
the measure $d\mu(Z,Z^{\dagger})$ that satisfies the resolution 
of unity 
\begin{equation}
 \label{eq:resolution-of-unity}
  \int_{S} d\mu(Z,Z^{\dagger}) \ket{Z}\bra{Z} = {\bf 1}_{V}\quad  
  \mbox{and}\quad \int_{S} d\mu(Z,Z^{\dagger}) = \mbox{dim}V\ .
\end{equation}

Next we define an anti-automorphism $\flat : S \longrightarrow S$. We call 
$Z \longrightarrow {Z}^{\flat}$ an anti-automorphism if and only if
\begin{eqnarray}
 \label{eq:automorphism} 
 &&(\mbox{i})\ \ Z \longrightarrow {Z}^{\flat}\  \mbox{induces an 
   automorphism of}\ S, \\
 \label{eq:anti-map} 
 &&(\mbox{ii})\ \ {\flat}\  \mbox{is an anti-map, namely}\quad
   \braket{Z^{\flat}}{W^{\flat}}=\braket{W}{Z}. 
\end{eqnarray}
Now let us define the generalized Bell state \cite{DF} :
\par \noindent
 \textbf{Definition} (Fivel)\  The generalized Bell state is defined as 
\begin{equation}
 \label{eq:generalized-Bell-state}
  \bell{B}= \frac{1}{\sqrt{\mbox{dim}V}} \int_{S} d\mu(Z,Z^{\dagger}) 
            \ket{Z}\otimes \ket{Z^{\flat}}.
\end{equation}
Then we have 
\begin{eqnarray}
  \bellbraket{B}{B}&=&\frac{1}{\mbox{dim}V} \int_{S} \int_{S} 
     d\mu(Z,Z^{\dagger})d\mu(W,W^{\dagger})  
     (\bra{Z}\otimes \bra{Z^{\flat}})( \ket{W}\otimes \ket{W^{\flat}})
  \nonumber \\
  &=&\frac{1}{\mbox{dim}V} \int_{S} \int_{S}
     d\mu(Z,Z^{\dagger})d\mu(W,W^{\dagger})
    \braket{Z}{W}\braket{Z^{\flat}}{W^{\flat}} 
  \nonumber \\
  &=&\frac{1}{\mbox{dim}V} \int_{S} \int_{S}
     d\mu(Z,Z^{\dagger})d\mu(W,W^{\dagger})
    \braket{Z}{W}\braket{W}{Z} 
  \nonumber \\
  &=&\frac{1}{\mbox{dim}V} \int_{S}
     d\mu(Z,Z^{\dagger})\braket{Z}{Z} 
   = \frac{1}{\mbox{dim}V} \int_{S} d\mu(Z,Z^{\dagger}) 
   = 1, \nonumber
\end{eqnarray}
where we have used (\ref{eq:resolution-of-unity}) and 
(\ref{eq:anti-map}). 

\par \noindent
Therefore we can get several generalized Bell states as choosing several 
anti-automorphisms. In the next section we will show that these states just 
coincide with the famous Bell states \cite{JB} 
in the case of spin $\frac{1}{2}$ : 
\begin{eqnarray}
  \label{eq:Bell-states}
 && \frac{1}{\sqrt{2}}
      (\ket{0}\otimes \ket{0} + \ket{1}\otimes \ket{1}), \quad 
    \frac{1}{\sqrt{2}}
      (\ket{0}\otimes \ket{0} - \ket{1}\otimes \ket{1}),  \nonumber \\
 && \frac{1}{\sqrt{2}}
      (\ket{0}\otimes \ket{1} + \ket{1}\otimes \ket{0}), \quad 
    \frac{1}{\sqrt{2}}
      (\ket{0}\otimes \ket{1} - \ket{1}\otimes \ket{0}).
\end {eqnarray}

\vspace{5mm}
Next we make a review of complex projective spaces, \cite{MN}, \cite{FKSF2} 
and \cite{KF5}. 
For $N \in \futon$ the complex projective space ${\fukuso}P^{N}$ is defined 
as follows : For \mbox{\boldmath $\zeta$}, \mbox{\boldmath $\mu$} $\in 
{\fukuso}^{N+1}-\{{\bf 0}\}$\   \mbox{\boldmath $\zeta$} is equivalent to 
\mbox{\boldmath $\mu$} (\mbox{\boldmath $\zeta$} $\sim$ 
\mbox{\boldmath $\mu$}) if and only if \mbox{\boldmath $\zeta$} = $\lambda$
\mbox{\boldmath $\mu$} for some $\lambda \in \fukuso - \{0 \}$. We show 
its equivalence relation class as [\mbox{\boldmath $\zeta$}] and set 
${\fukuso}P^{N} \equiv {\fukuso}^{N+1}-\{{\bf 0}\} / \sim $. When  
\mbox{\boldmath $\zeta$} = $({\zeta}_{0}, {\zeta}_{1}, \cdots, {\zeta}_{N})$ 
we write usually as [\mbox{\boldmath $\zeta$}] = $[{\zeta}_{0}: {\zeta}_{1}:  
\cdots : {\zeta}_{N}]$. Then it is well--known that ${\fukuso}P^{N}$ has 
$N+1$ local charts, namely
\begin{equation}
  {\fukuso}P^{N} = \bigcup_{j=0}^{N} U_{j}\ ,  \quad 
    U_{j} = \{ [{\zeta}_{0}: \cdots : {\zeta}_{j}: \cdots : {\zeta}_{N}]\ |\  
          {\zeta}_{j} \ne 0 \}.
\end{equation} 
Since
\[
  ({\zeta}_{0}, \cdots , {\zeta}_{j}, \cdots , {\zeta}_{N}) =  
  {\zeta}_{j}\left(\frac{{\zeta}_{0}}{{\zeta}_{j}}, \cdots, 
  \frac{{\zeta}_{j-1}}{{\zeta}_{j}}, 1, \frac{{\zeta}_{j+1}}{{\zeta}_{j}}, 
  \cdots, \frac{{\zeta}_{N}}{{\zeta}_{j}}\right),
\]
we have the local coordinate on $U_{j}$ 
\begin{equation}
  \left(\frac{{\zeta}_{0}}{{\zeta}_{j}}, \cdots, 
  \frac{{\zeta}_{j-1}}{{\zeta}_{j}}, \frac{{\zeta}_{j+1}}{{\zeta}_{j}}, 
  \cdots, \frac{{\zeta}_{N}}{{\zeta}_{j}}\right). 
\end{equation}

But the above definition of ${\fukuso}P^{N}$ is not handy, so we use 
the well--known expression by projections
\begin{equation}
 {\fukuso}P^{N} \cong G_{1}({\fukuso}^{N+1}) = 
     \{P \in M(N+1; \fukuso)\ |\ P^{2} = P,\ P^{\dagger} = P \ \mbox{and}\ 
       \mbox{tr}P = 1 \}
\end{equation}
and this correspondence 
\begin{equation}
 \label{eq:correspondence}
  [{\zeta}_{0}: {\zeta}_{1}: \cdots : {\zeta}_{N}] \Longleftrightarrow 
  \frac{1}{\zetta{{\zeta}_{0}}^2 + \zetta{{\zeta}_{1}}^2 + \cdots + 
          \zetta{{\zeta}_{N}}^2 }
  \left(
     \begin{array}{ccccc} 
         \zetta{{\zeta}_{0}}^2& {\zeta}_{0}{\bar {\zeta}_{1}}& 
         \cdot& \cdot& {\zeta}_{0}{\bar {\zeta}_{N}}  \\
         {\zeta}_{1}{\bar {\zeta}_{0}} & \zetta{{\zeta}_{1}}^2&  
         \cdot& \cdot& {\zeta}_{1}{\bar {\zeta}_{N}}  \\
         \cdot& \cdot& & & \cdot  \\
         \cdot& \cdot& & & \cdot  \\
         {\zeta}_{N}{\bar {\zeta}_{0}}& {\zeta}_{N}{\bar {\zeta}_{1}}& 
         \cdot& \cdot& \zetta{{\zeta}_{N}}^2     
     \end{array}
  \right) \equiv P\ .
\end{equation}
If we set 
\begin{equation}
  \ket{\mbox{\boldmath $\zeta$}}=
 \frac{1}{\sqrt{\sum_{j=0}^{N} \zetta{\zeta_{j}}^2} }
  \left(
     \begin{array}{c}
        {\zeta}_{0} \\
        {\zeta}_{1} \\
         \cdot  \\
         \cdot  \\
        {\zeta}_{N} 
     \end{array} 
  \right)\ , 
\end{equation}
then we can write the right hand side of (\ref{eq:correspondence}) as 
\begin{equation}
 \label{eq:projector}
  P = \ket{\mbox{\boldmath $\zeta$}}\bra{\mbox{\boldmath $\zeta$}} \quad 
  \mbox{and} \quad 
    \braket{\mbox{\boldmath $\zeta$}}{\mbox{\boldmath $\zeta$}} = 1.
\end{equation}
For example on $U_{1}$ 
\[
  \left(z_{1}, z_{2}, \cdots, z_{N} \right) = 
  \left(\frac{{\zeta}_{1}}{{\zeta}_{0}},\frac{{\zeta}_{2}}{{\zeta}_{0}}, 
  \cdots, 
  \frac{{\zeta}_{N}}{{\zeta}_{0}}\right) ,  
\]
we have 
\begin{eqnarray}
  P(z_{1}, \cdots, z_{N}) &=& 
  \frac{1}{1 + \sum_{j=1}^{N} \zetta{z_{j}}^2}
     \left(
         \begin{array}{ccccc}
             1& {\bar z_{1}}& \cdot& \cdot& {\bar z_{N}}  \\
             z_{1}& \zetta{z_{1}}^2& \cdot& \cdot& z_{1}{\bar z_{N}} \\
             \cdot& \cdot& & & \cdot \\
             \cdot& \cdot& & & \cdot \\
             z_{N}& z_{N}{\bar z_{1}}& \cdot& \cdot& \zetta{z_{N}}^2
         \end{array}
     \right)  \nonumber \\
   &=& \ket{\left(z_{1}, z_{2}, \cdots, z_{N}\right)}
       \bra{\left(z_{1}, z_{2}, \cdots, z_{N}\right)}\ ,
\end{eqnarray}
where 
\begin{equation}
  \ket{\left(z_{1}, z_{2}, \cdots, z_{N} \right)} = 
  \frac{1}{\sqrt{1 + \sum_{j=1}^{N} \zetta{z_{j}}^2}}
     \left(
         \begin{array}{c}
             1 \\
             z_{1} \\
             \cdot \\
             \cdot \\
             z_{N} 
         \end{array}
     \right).  \nonumber \\
\end{equation}

For the latter use let us give a more detail description for the 
cases $N$ = $1$ and $2$.

\par \noindent
 (a) {\bf $N = 1$} : 
\begin{eqnarray}
 \label{eq:cp1-1}
  P(z)&=&\frac{1}{1+\zetta{z}^2}
     \left(
         \begin{array}{cc}
             1& {\bar z} \\
             z& \zetta{z}^2 
         \end{array}
     \right)  
   = \ket{z}\bra{z}, \nonumber  \\
  &&\mbox{where}\ \ket{z}=\frac{1}{\sqrt{1+\zetta{z}^2}}
     \left(
         \begin{array}{c}
             1 \\
             z 
         \end{array}
     \right), 
  \quad z=\frac{\zeta_{1}}{\zeta_{0}}, \quad \mbox{on}\ U_{1}\ ,  \\
 \label{eq:cp1-2}
  P(w)&=&\frac{1}{\zetta{w}^2+1}
     \left(
         \begin{array}{cc}
             \zetta{w}^2 & w \\
             {\bar w}& 1
         \end{array}
     \right)  
   = \ket{w}\bra{w},  \nonumber  \\
  &&\mbox{where}\ \ket{w}=\frac{1}{\sqrt{\zetta{w}^2+1}}
     \left(
         \begin{array}{c}
             w \\
             1
         \end{array}
     \right), 
  \quad w=\frac{\zeta_{0}}{\zeta_{1}}, \quad \mbox{on}\ U_{2}\ . 
\end{eqnarray}

\vspace{5mm}
\par \noindent
 (b) {\bf $N = 2$} : 
\begin{eqnarray}
 \label{eq:cp2-1}
  P(z_{1},z_{2})&=&\frac{1}{1+\zetta{z_{1}}^2+\zetta{z_{2}}^2}
     \left(
         \begin{array}{ccc}
             1& {\bar z_{1}}& {\bar z_{2}} \\
             z_{1}& \zetta{z_{1}}^2& z_{1}{\bar z_{2}} \\
             z_{2}& z_{2}{\bar z_{1}}& \zetta{z_{2}}^2 
         \end{array}
     \right)  
   = \ket{(z_{1},z_{2})}\bra{(z_{1},z_{2})}, \nonumber  \\
  \mbox{where} 
  &&\ket{(z_{1},z_{2})}=\frac{1}{\sqrt{1+\zetta{z_{1}}^2+\zetta{z_{2}}^2}}
     \left(
         \begin{array}{c}
             1 \\
             z_{1} \\
             z_{2} 
         \end{array}
     \right), 
\quad (z_{1},z_{2})=\left(\frac{\zeta_{1}}{\zeta_{0}},
   \frac{\zeta_{2}}{\zeta_{0}} \right)  \quad \mbox{on}\ U_{1}\ ,  \\
 \label{eq:cp2-2}
  P(w_{1},w_{2})&=&\frac{1}{\zetta{w_{1}}^2+1+\zetta{w_{2}}^2}
     \left(
         \begin{array}{ccc}
             \zetta{w_{1}}^2& w_{1}& w_{1}{\bar w_{2}} \\
             {\bar w_{1}}& 1& {\bar w_{2}} \\
             w_{2}{\bar w_{1}}& w_{2}& \zetta{w_{2}}^2 
         \end{array}
     \right)  
   = \ket{(w_{1},w_{2})}\bra{(w_{1},w_{2})}, \nonumber \\
  \mbox{where} 
  &&\ket{(w_{1},w_{2})}=\frac{1}{\sqrt{\zetta{w_{1}}^2+1+\zetta{w_{2}}^2}}
     \left(
         \begin{array}{c}
             w_{1} \\
              1  \\
             w_{2} 
         \end{array}
     \right), 
\quad  (w_{1},w_{2})=\left(\frac{\zeta_{0}}{\zeta_{1}},
   \frac{\zeta_{2}}{\zeta_{1}} \right)\ \  \mbox{on}\ U_{2}\ , \\
 \label{eq:cp2-3}
  P(v_{1},v_{2})&=&\frac{1}{\zetta{v_{1}}^2+\zetta{v_{2}}^2+1}
     \left(
         \begin{array}{ccc}
             \zetta{v_{1}}^2& v_{1}{\bar v_{2}}& v_{1} \\
             v_{2}{\bar v_{1}}& \zetta{v_{2}}^2& v_{2} \\
             {\bar v_{1}}& {\bar v_{2}}& 1 
         \end{array}
     \right)  
   = \ket{(v_{1},v_{2})}\bra{(v_{1},v_{2})}, \nonumber \\
  \mbox{where} 
  &&\ket{(v_{1},v_{2})}=\frac{1}{\sqrt{\zetta{v_{1}}^2+\zetta{v_{2}}^2+1}}
     \left(
         \begin{array}{c}
             v_{1} \\
             v_{2}  \\
              1 
         \end{array}
     \right), 
\quad (v_{1},v_{2})=\left(\frac{\zeta_{0}}{\zeta_{2}},
   \frac{\zeta_{1}}{\zeta_{2}} \right)  \quad \mbox{on}\ U_{3}\ . 
\end{eqnarray}

\vspace{1cm}
\section{Bell States Revisited}
In this section we show that (\ref{eq:generalized-Bell-state}) coinsides with 
the Bell states (\ref{eq:Bell-states}) by choosing anti-automorphism $\flat$ 
suitably.

First let us recall the spin $j$--representation of Lie algebra $su(2)$ from 
\cite{FKSF1}. This is a coherent representation of $su(2)$ based on complex 
manifold ${\fukuso}P^{1}$ in our terminology. The algebra of $\{J_{+},J_{-},
J_{3}\}$ reads  
\begin{equation} 
  [J_{3},J_{+}]=J_{+},\ [J_{3},J_{-}]=-J_{-},\ [J_{+},J_{-}]=2J_{3}\ ,
\end{equation}
where $J_{\pm}=\frac{1}{2}(J_{1}\pm iJ_{2})$ and actions of $\{J_{+},J_{-},
J_{3}\}$ on a representation space $V$ ($\cong {\fukuso}^{2j+1}$) 
\begin{eqnarray} 
    J_{+}\ket{{j,m}}&=&\sqrt{(j-m)(j+m+1)}\ket{{j,m+1}},\  
    J_{-}\ket{{j,m}}=\sqrt{(j-m+1)(j+m)}\ket{{j,m-1}}, \nonumber \\
    J_{3}\ket{{j,m}}&=& m\ket{{j,m}}, 
\end{eqnarray}
where $-j \le m \le j$. We note 
\begin{equation} 
   {\bf 1}_{j}=\sum_{m=-j}^{j}\ket{{j,m}}\bra{{j,m}}\quad \mbox{and}\quad 
   \braket{{j,m}}{{j,n}}=\delta_{mn}.
\end{equation}
Then the coherent state $\ket{z}$ ($z \in \fukuso \subset {\fukuso}P^1$) is 
defined as
\begin{equation}
  \ket{z}=\frac{1}{\left(1+\zetta{z}^2 \right)^j}
    \sum_{k=0}^{2j}\sqrt{{}_{2j}C_k}z^k \ket{{j,-j+k}}  
\end{equation}
and this satisfies the resolution of unity (\ref{eq:resolution-of-unity}) 
\begin{equation}
 \int_{\fukuso}d\mu(z,{\bar z})\ket{z}\bra{z}= 
     \sum_{m=-j}^{j}\ket{{j,m}}\bra{{j,m}}={\bf 1}_{j}
 \quad \mbox{and} \quad 
 \int_{\fukuso}d\mu(z,{\bar z})=2j+1\ ,
\end{equation}
where the measure $d\mu(z,{\bar z})$ is 
\begin{equation}
  d\mu(z,{\bar z})=\frac{2j+1}{\pi}\frac{[d^{2}z]}{(1+\zetta{z}^2)^2}\ .
\end{equation}
We note that this measure is invariant under the transform $z \longrightarrow  
1/z$, so this one is defined on ${\fukuso}P^1$ not $\fukuso$. 
In the following we set for simplicity
\begin{equation}
      \ket{{j,-j+k}}=\ket{k} \quad \mbox{for}\quad  0 \le k \le 2j\ . 
\end{equation}
For example\ $\ket{{\frac{1}{2},-\frac{1}{2}}}=\ket{0}$ and 
$\ket{{\frac{1}{2},\frac{1}{2}}}=\ket{1}$\  in the case of spin $\frac{1}{2}$.
In this case we consider the following four anti-automorphisms 
(\ref{eq:automorphism}) and (\ref{eq:anti-map})
:
\begin{equation}
 \label{eq:4-anti-automorphisms}
  \mbox{(1)}\ \  z^{\flat}={\bar z}\qquad  
  \mbox{(2)}\ \  z^{\flat}=-{\bar z}\qquad  
  \mbox{(3)}\ \  z^{\flat}=\frac{1}{\bar z}\qquad  
  \mbox{(4)}\ \  z^{\flat}=\frac{-1}{\bar z}\ .
\end{equation}
Then it is easy to see from (\ref{eq:cp1-1}) and (\ref{eq:cp1-2}) 
\vspace{5mm}
\par \noindent 
\begin{Large}
 \textbf{Lemma 1}
\end{Large}
\begin{eqnarray}
&&\mbox{(1)}\ \ \ket{z^{\flat}}=\ket{{\bar z}}=
      \frac{1}{\sqrt{1+\zetta{z}^2}}(\ket{0}+{\bar z}\ket{1}), \\
&&\mbox{(2)}\ \ \ket{z^{\flat}}=\ket{-{\bar z}}=
      \frac{1}{\sqrt{1+\zetta{z}^2}}(\ket{0}-{\bar z}\ket{1}), \\
&&\mbox{(3)}\ \ \ket{z^{\flat}}=\ket{1/{\bar z}}=
      \frac{1}{\sqrt{1+\zetta{z}^2}}({\bar z}\ket{0}+\ket{1}), \\
&&\mbox{(4)}\ \ \ket{z^{\flat}}=\ket{-1/{\bar z}}=
      \frac{1}{\sqrt{1+\zetta{z}^2}}({\bar z}\ket{0}-\ket{1}).
\end{eqnarray}
Here we have identified 
$\ket{0}= {1\choose 0}$ and $\ket{1}={0\choose 1}$. 
Then making use of elementary facts  
\begin{eqnarray}
&&\frac{2}{\pi}\int_{\fukuso} \frac{[d^{2}z]}{(1+\zetta{z}^2)^2}
   \frac{1}{1+\zetta{z}^2}=
 \frac{2}{\pi}\int_{\fukuso} \frac{[d^{2}z]}{(1+\zetta{z}^2)^2}
   \frac{\zetta{z}^2}{1+\zetta{z}^2}=1, \nonumber \\
&&\frac{2}{\pi}\int_{\fukuso} \frac{[d^{2}z]}{(1+\zetta{z}^2)^2}
   \frac{z}{1+\zetta{z}^2}=
 \frac{2}{\pi}\int_{\fukuso} \frac{[d^{2}z]}{(1+\zetta{z}^2)^2}
   \frac{{\bar z}}{1+\zetta{z}^2}=0, \nonumber 
\end{eqnarray}
we have easily 
\vspace{5mm}
\par \noindent 
\begin{Large}
 \textbf{Proposition 2}
\end{Large}
\begin{eqnarray}
&&\mbox{(1)}\ \ \bell{B}=\frac{1}{\sqrt{2}}\int_{\fukuso} d\mu(z,{\bar z})
  \ket{z}\otimes \ket{{\bar z}}
 = \frac{1}{\sqrt{2}}(\ket{0}\otimes \ket{0} + \ket{1}\otimes \ket{1}), \\
&&\mbox{(2)}\ \ \bell{B}=\frac{1}{\sqrt{2}}\int_{\fukuso} d\mu(z,{\bar z})
  \ket{z}\otimes \ket{-{\bar z}}
 = \frac{1}{\sqrt{2}}(\ket{0}\otimes \ket{0} - \ket{1}\otimes \ket{1}), \\
&&\mbox{(3)}\ \ \bell{B}=\frac{1}{\sqrt{2}}\int_{\fukuso} d\mu(z,{\bar z})
  \ket{z}\otimes \ket{1/{\bar z}}
 = \frac{1}{\sqrt{2}}(\ket{0}\otimes \ket{1} + \ket{1}\otimes \ket{0}), \\
&&\mbox{(4)}\ \ \bell{B}=\frac{1}{\sqrt{2}}\int_{\fukuso} d\mu(z,{\bar z})
  \ket{z}\otimes \ket{-1/{\bar z}}
 = \frac{1}{\sqrt{2}}(\ket{0}\otimes \ket{1} - \ket{1}\otimes \ket{0}), 
\end{eqnarray}
where $d\mu(z,{\bar z})=\frac{2}{\pi}\frac{[d^{2}z]}{(1+\zetta{z}^2)^2}$. 
We note that our calculation is based on the following two matrices :
\begin{equation}
 \label{eq:sigma31}
  \sigma_{3}=\left(
     \begin{array}{cc}
        1&  \\
         &-1
      \end{array}
    \right), \quad
  \sigma_{1}=\left(
     \begin{array}{cc}
         & 1  \\
        1 & 
      \end{array}
    \right)\ .
\end{equation}
We just recovered the Bell states (\ref{eq:Bell-states}) !! \ \ 
The author does not know whether this result has been known or not. 

Since we consider the case of spin $j$, we write $\ket{z}$ as 
\begin{equation}
  {\ket{z}}_{j}=\frac{1}{\left(1+\zetta{z}^2 \right)^j}
    \sum_{k=0}^{2j}\sqrt{{}_{2j}C_k}z^k \ket{k}  
\end{equation}
to emphasize the dependence of spin $j$. From Proposition 2 it is very 
natural to define Bell states with spin $j$ as follows because the 
parameter space is the same ${\fukuso}P^1$ : 
\vspace{5mm}
\par \noindent 
\begin{Large}
 \textbf{Definition 3} (Bell states with spin $j$)
\end{Large}
\begin{eqnarray}
&&\mbox{(1)}\ \ \bell{B}=\frac{1}{\sqrt{2j+1}}\int_{\fukuso} d\mu(z,{\bar z})
  {\ket{z}}_{j}\otimes {\ket{{\bar z}}}_{j},  \\
&&\mbox{(2)}\ \ \bell{B}=\frac{1}{\sqrt{2j+1}}\int_{\fukuso} d\mu(z,{\bar z})
  {\ket{z}}_{j}\otimes {\ket{-{\bar z}}}_{j}, \\
&&\mbox{(3)}\ \ \bell{B}=\frac{1}{\sqrt{2j+1}}\int_{\fukuso} d\mu(z,{\bar z})
  {\ket{z}}_{j}\otimes {\ket{1/{\bar z}}}_{j}, \\
&&\mbox{(4)}\ \ \bell{B}=\frac{1}{\sqrt{2j+1}}\int_{\fukuso} d\mu(z,{\bar z})
  {\ket{z}}_{j}\otimes {\ket{-1/{\bar z}}}_{j}, 
\end{eqnarray}
where $d\mu(z,{\bar z})=\frac{2j+1}{\pi}\frac{[d^{2}z]}{(1+\zetta{z}^2)^2}$.
Let us calculate ${\ket{{\bar z}}}_{j}$, ${\ket{-{\bar z}}}_{j}$, 
${\ket{1/{\bar z}}}_{j}$ and ${\ket{-1/{\bar z}}}_{j}$. It is easy to see 
\vspace{5mm}
\par \noindent 
\begin{Large}
 \textbf{Lemma 4}
\end{Large}
\begin{eqnarray}
&&\mbox{(1)}\ \ {\ket{{\bar z}}}_{j}=\frac{1}{\left(1+\zetta{z}^2 \right)^j}
    \sum_{k=0}^{2j}\sqrt{{}_{2j}C_k}{\bar z}^k \ket{k}, \\  
&&\mbox{(2)}\ \ {\ket{-{\bar z}}}_{j}=\frac{1}{\left(1+\zetta{z}^2 \right)^j}
    \sum_{k=0}^{2j}\sqrt{{}_{2j}C_k}(-1)^k{\bar z}^k \ket{k}, \\  
&&\mbox{(3)}\ \ {\ket{1/{\bar z}}}_{j}=
    \frac{1}{\left(1+\zetta{z}^2 \right)^j}
    \sum_{k=0}^{2j}\sqrt{{}_{2j}C_k}{\bar z}^k \ket{2j-k}, \\  
&&\mbox{(4)}\ \ {\ket{-1/{\bar z}}}_{j}=
    \frac{1}{\left(1+\zetta{z}^2 \right)^j}
    \sum_{k=0}^{2j}\sqrt{{}_{2j}C_k}(-1)^k{\bar z}^k \ket{2j-k}. 
\end{eqnarray}
From this lemma and the elementary facts  
\[
  \frac{2j+1}{\pi}\int_{\fukuso} \frac{[d^{2}z]}{(1+\zetta{z}^2)^2}
   \frac{\zetta{z}^{2k}}{(1+\zetta{z}^2)^{2j}}=\frac{1}{{}_{2j}C_k}
  \quad \mbox{for}\quad 0 \le k \le 2j\ ,
\]
we can give explicit forms to the Bell states with spin $j$ 
in Definition 3 : 
\vspace{5mm}
\par \noindent 
\begin{Large}
 \textbf{Proposition 5}
\end{Large}
\begin{eqnarray}
&&\mbox{(1)}\ \ \bell{B}=\frac{1}{\sqrt{2j+1}} \sum_{k=0}^{2j}
   \ket{k}\otimes \ket{k},  \\
&&\mbox{(2)}\ \ \bell{B}=\frac{1}{\sqrt{2j+1}} \sum_{k=0}^{2j}(-1)^{k}
   \ket{k}\otimes \ket{k}, \\
&&\mbox{(3)}\ \ \bell{B}=\frac{1}{\sqrt{2j+1}} \sum_{k=0}^{2j}
  \ket{k}\otimes \ket{2j-k}, \\
&&\mbox{(4)}\ \ \bell{B}=\frac{1}{\sqrt{2j+1}} \sum_{k=0}^{2j}(-1)^{k}
  \ket{k}\otimes \ket{2j-k}.
\end{eqnarray}

\par \noindent 
A comment is in order. We list the above result once more for the case $j=1$ :
\begin{eqnarray}
&&\mbox{(1)}\ \ \frac{1}{\sqrt{3}} 
   (\ket{0}\otimes \ket{0}+\ket{1}\otimes \ket{1}+\ket{2}\otimes \ket{2}), 
   \nonumber \\
&&\mbox{(2)}\ \ \frac{1}{\sqrt{3}} 
   (\ket{0}\otimes \ket{0}-\ket{1}\otimes \ket{1}+\ket{2}\otimes \ket{2}), 
  \nonumber \\
&&\mbox{(3)}\ \ \frac{1}{\sqrt{3}} 
   (\ket{0}\otimes \ket{2}+\ket{1}\otimes \ket{1}+\ket{2}\otimes \ket{0}), 
   \nonumber \\
&&\mbox{(4)}\ \ \frac{1}{\sqrt{3}} 
   (\ket{0}\otimes \ket{2}-\ket{1}\otimes \ket{1}+\ket{2}\otimes \ket{0}).
    \nonumber 
\end{eqnarray}
It is easy to see that they are not linearly independent. Namely this case 
is very special (peculiar).

\vspace{1cm}
\section{Generalized Bell States}
In this section we generalize the result in the preceeding section, namely 
we treat the coherent states of $u(n+1)$ based on ${\fukuso}P^n$ (see 
\cite{FKSF2})  
and calculate generalized Bell states (\ref{eq:generalized-Bell-state}) 
for several anti-automorphisms like (\ref{eq:4-anti-automorphisms}). 
But to avoid complicated calculations we consider the case $n=2$ and $Q=1$ 
only, because it is easy to conjecture the corresponding result in general 
case from this special case.

Let $\{\ket{0},\ket{1},\ket{2}\}$ be a basis of the representation space 
$V$ ($\cong {\fukuso}^3$). Namely 
\[
 \sum_{j=0}^{2}\ket{j}\bra{j}={\bf 1}_{Q=1}
 \quad \mbox{and} \quad  \braket{i}{j}=\delta_{ij}.
\]
A coherent state $\ket{(z_1,z_2)}$ for $(z_1,z_2)\in {\fukuso}^2$ 
is defined as
\begin{equation}
 \ket{(z_1,z_2)}=\frac{1}{\sqrt{1+\zetta{z_1}^2+\zetta{z_2}^2}}
  (\ket{0}+z_1\ket{1}+z_2\ket{2})
\end{equation}
and the measure 
\begin{equation}
  d\mu(Z,Z^{\dagger})=\frac{6}{\pi^2}
    \frac{[d^2z_1][d^2z_2]}{(1+\zetta{z_1}^2+\zetta{z_2}^2)^3}.
\end{equation}
Then we have 
\begin{equation}
 \int_{\fukuso^2}d\mu(Z,Z^{\dagger}) \ket{(z_1,z_2)}\bra{(z_1,z_2)}
  =\sum_{j=0}^{2}\ket{j}\bra{j}={\bf 1}_{Q=1} 
 \quad \mbox{and}\quad 
 \int_{\fukuso^2}d\mu(Z,Z^{\dagger})=3\ .
\end{equation}

Let $\omega$ be an element in $\fukuso$ satisfying ${\omega}^3=1$. 
Then $1+\omega+{\omega}^2=0$ and ${\bar \omega}={\omega}^2$.\ \ 
Here we consider the following nine anti-automorphisms 
(\ref{eq:automorphism}) and (\ref{eq:anti-map}) :
\begin{eqnarray}
 \label{eq:9-a-anti-automorphisms}
  &&\mbox{(a--1, 2, 3)}\quad
  (z_{1},z_{2})^{\flat}=({\bar z_{1}},{\bar z_{2}}), \  
  (z_{1},z_{2})^{\flat}=({\omega}{\bar z_{1}},{\omega}^2{\bar z_{2}}), \ 
  (z_{1},z_{2})^{\flat}=({\omega}^2{\bar z_{1}},{\omega}{\bar z_{2}}), \\
 \label{eq:9-b-anti-automorphisms}
 &&\mbox{(b--1, 2, 3)}\quad   
   (z_{1},z_{2})^{\flat}=
  (\frac{1}{{\bar z_{2}}},\frac{{\bar z_{1}}}{\bar z_{2}}), \ 
  (z_{1},z_{2})^{\flat}=
  (\omega\frac{1}{{\bar z_{2}}},{\omega}^2\frac{{\bar z_{1}}}{\bar z_{2}}), \ 
  (z_{1},z_{2})^{\flat}=
  ({\omega}^2\frac{1}{{\bar z_{2}}},\omega\frac{{\bar z_{1}}}{\bar z_{2}}), \\ 
 \label{eq:9-c-anti-automorphisms}
  &&\mbox{(c--1, 2, 3)}\quad   
  (z_{1},z_{2})^{\flat}=
  (\frac{{\bar z_{2}}}{\bar z_{1}},\frac{1}{{\bar z_{1}}}), \ 
  (z_{1},z_{2})^{\flat}=
  (\omega\frac{{\bar z_{2}}}{\bar z_{1}},{\omega}^2\frac{1}{{\bar z_{1}}}), \ 
  (z_{1},z_{2})^{\flat}= 
  ({\omega}^2\frac{{\bar z_{2}}}{\bar z_{1}},\omega\frac{1}{{\bar z_{1}}}).
\end{eqnarray}
A note is in order. We can of course choose another anti-automorphisms instead 
of the above ones.  
\par \noindent 
Then it is easy to see from (\ref{eq:cp2-1}), (\ref{eq:cp2-2}) and 
(\ref{eq:cp2-3})
\vspace{3mm}
\par \noindent 
\begin{Large}
 \textbf{Lemma 6}
\end{Large}
\begin{eqnarray}
  &&\mbox{(a--1)}\ 
 \ket{(z_{1},z_{2})^{\flat}}=
 \ket{({\bar z_{1}},{\bar z_{2}})}=
 \frac{1}{\sqrt{1+\zetta{z_1}^2+\zetta{z_2}^2}}
 (\ket{0}+{\bar z_1}\ket{1}+{\bar z_2}\ket{2}), \\
  &&\mbox{(a--2)}\ 
 \ket{(z_{1},z_{2})^{\flat}}=
 \ket{(\omega{\bar z_{1}},\omega^2{\bar z_{2}})}=
 \frac{1}{\sqrt{1+\zetta{z_1}^2+\zetta{z_2}^2}}
 (\ket{0}+\omega{\bar z_1}\ket{1}+\omega^2{\bar z_2}\ket{2}), \\
  &&\mbox{(a--3)}\ 
 \ket{(z_{1},z_{2})^{\flat}}=
 \ket{(\omega^2{\bar z_{1}},\omega{\bar z_{2}})}=
 \frac{1}{\sqrt{1+\zetta{z_1}^2+\zetta{z_2}^2}}
 (\ket{0}+\omega^2{\bar z_1}\ket{1}+\omega{\bar z_2}\ket{2}), \\
  &&\mbox{(b--1)}\ 
 \ket{(z_{1},z_{2})^{\flat}}=
 \ket{(1/{\bar z_2},{\bar z_1}/{\bar z_2})}=
 \frac{1}{\sqrt{1+\zetta{z_1}^2+\zetta{z_2}^2}}
 ({\bar z_2}\ket{0}+\ket{1}+{\bar z_1}\ket{2}), \\
  &&\mbox{(b--2)}\ 
 \ket{(z_{1},z_{2})^{\flat}}=
 \ket{(\omega/{\bar z_2},\omega^2{\bar z_1}/{\bar z_2})}=
 \frac{1}{\sqrt{1+\zetta{z_1}^2+\zetta{z_2}^2}}
 (\omega^2{\bar z_2}\ket{0}+\ket{1}+\omega{\bar z_1}\ket{2}), \\
  &&\mbox{(b--3)}\ 
 \ket{(z_{1},z_{2})^{\flat}}=
 \ket{(\omega^2/{\bar z_2},\omega{\bar z_1}/{\bar z_2})}=
 \frac{1}{\sqrt{1+\zetta{z_1}^2+\zetta{z_2}^2}}
 (\omega{\bar z_2}\ket{0}+\ket{1}+\omega^2{\bar z_1}\ket{2}), \\
  &&\mbox{(c--1)}\ 
 \ket{(z_{1},z_{2})^{\flat}}=
 \ket{({\bar z_2}/{\bar z_1},1/{\bar z_1})}=
 \frac{1}{\sqrt{1+\zetta{z_1}^2+\zetta{z_2}^2}}
 ({\bar z_1}\ket{0}+{\bar z_2}\ket{1}+\ket{2}), \\
  &&\mbox{(c--2)}\ 
 \ket{(z_{1},z_{2})^{\flat}}=
 \ket{(\omega{\bar z_2}/{\bar z_1},\omega^2/{\bar z_1})}=
 \frac{1}{\sqrt{1+\zetta{z_1}^2+\zetta{z_2}^2}}
 (\omega{\bar z_1}\ket{0}+\omega^2{\bar z_2}\ket{1}+\ket{2}), \\
  &&\mbox{(c--3)}\ 
 \ket{(z_{1},z_{2})^{\flat}}=
 \ket{(\omega^2{\bar z_2}/{\bar z_1},\omega/{\bar z_1})}=
 \frac{1}{\sqrt{1+\zetta{z_1}^2+\zetta{z_2}^2}}
 (\omega^2{\bar z_1}\ket{0}+\omega{\bar z_2}\ket{1}+\ket{2}).
\end{eqnarray}
\vspace{2mm}
\par \noindent
From this lemma and the elementary facts  
\begin{eqnarray}
  &&\frac{6}{\pi^2}\int_{\fukuso^2} \frac{[d^{2}z_1][d^{2}z_2]}
{(1+\zetta{z_1}^2+\zetta{z_2}^2)^3}
   \frac{1}{1+\zetta{z_1}^2+\zetta{z_2}^2}=
  \frac{6}{\pi^2}\int_{\fukuso^2} \frac{[d^{2}z_1][d^{2}z_2]}
{(1+\zetta{z_1}^2+\zetta{z_2}^2)^3}
   \frac{\zetta{z_1}^2}{1+\zetta{z_1}^2+\zetta{z_2}^2}=   \nonumber \\
  &&\frac{6}{\pi^2}\int_{\fukuso^2} \frac{[d^{2}z_1][d^{2}z_2]}
{(1+\zetta{z_1}^2+\zetta{z_2}^2)^3}
   \frac{\zetta{z_2}^2}{1+\zetta{z_1}^2+\zetta{z_2}^2}=1,  \nonumber 
\end{eqnarray}
we have easily 
\vspace{5mm}
\par \noindent 
\begin{Large}
 \textbf{Proposition 7} (generalized Bell states)
\end{Large}
\begin{eqnarray}
  &&\mbox{(a--1)}\ \  \bell{B}=\frac{1}{\sqrt{3}}
  (\ket{0}\otimes \ket{0}+\ket{1}\otimes \ket{1}+\ket{2}\otimes \ket{2}), \\
  &&\mbox{(a--2)}\ \  \bell{B}=\frac{1}{\sqrt{3}}
  (\ket{0}\otimes \ket{0}+\omega\ket{1}\otimes \ket{1}+
   \omega^2\ket{2}\otimes \ket{2}), \\
  &&\mbox{(a--3)}\ \  \bell{B}=\frac{1}{\sqrt{3}}
  (\ket{0}\otimes \ket{0}+\omega^2\ket{1}\otimes \ket{1}+
   \omega\ket{2}\otimes \ket{2}), \\
  &&\mbox{(b--1)}\ \  \bell{B}=\frac{1}{\sqrt{3}}
  (\ket{0}\otimes \ket{1}+\ket{1}\otimes \ket{2}+\ket{2}\otimes \ket{0}), \\
  &&\mbox{(b--2)}\ \  \bell{B}=\frac{1}{\sqrt{3}}
  (\ket{0}\otimes \ket{1}+\omega\ket{1}\otimes \ket{2}+
   \omega^2\ket{2}\otimes \ket{0}), \\
  &&\mbox{(b--3)}\ \  \bell{B}=\frac{1}{\sqrt{3}}
  (\ket{0}\otimes \ket{1}+\omega^2\ket{1}\otimes \ket{2}+
   \omega\ket{2}\otimes \ket{0}), \\
  &&\mbox{(c--1)}\ \  \bell{B}=\frac{1}{\sqrt{3}} 
  (\ket{0}\otimes \ket{2}+\ket{1}\otimes \ket{0}+\ket{2}\otimes \ket{1}), \\
  &&\mbox{(c--2)}\ \  \bell{B}=\frac{1}{\sqrt{3}}
  (\ket{0}\otimes \ket{2}+\omega\ket{1}\otimes \ket{0}+
   \omega^2\ket{2}\otimes \ket{1}), \\
  &&\mbox{(c--3)}\ \  \bell{B}=\frac{1}{\sqrt{3}}
  (\ket{0}\otimes \ket{2}+\omega^2\ket{1}\otimes \ket{0}+
   \omega\ket{2}\otimes \ket{1}).
\end{eqnarray}

\par \noindent
This is our main result. 
We note that our calculation is deeply based on the following two 
matrices :
\begin{equation}
 \label{eq:AB}
  A=\left(
     \begin{array}{ccc}
        1& &  \\
         &\omega &  \\
         & &\omega^2
      \end{array}
    \right), \quad
  B=\left(
     \begin{array}{ccc}
        0&  &1  \\
        1&0 &   \\
         &1 &0
      \end{array}
    \right).
\end{equation}
From Proposition 7 it is very easy to conjecture the explicit forms  
of generalized Bell states, \cite{KF6}.

\vspace{1cm}
\section{Discussion}
In this paper we calculated the generalized Bell states defined by Fivel,  
which are defined as the integral of tensor product of 
generalized coherent states based on ${\fukuso}P^N$ and their ``twisted''
ones due to anti--automorphisms on ${\fukuso}P^N$.   
The generalization to Grassmann manifolds $G_{k}({\fukuso}^{N+1})$ is under 
consideration, \cite{KF6}. 

But unfortunately the generalized Bell states by Fivel are not defined for 
coherent states based on non--compact complex manifolds such as the Poincare 
disk or Siegel domains. 
We need a drastic change of his definition. 

\vspace{1cm}
\noindent
{\it Acknowledgment.}
The author wishes to thank Yoshinori Machida for his warm hospitality 
at Numazu College of Technology. 
\vspace{1cm}
\par \noindent
\begin{Large}
  \textbf{Appendix A \ \ Anti--Automorphisms on $G_{1}({\fukuso}^{N+1})$}
\end{Large}
\vspace{3mm} \par \noindent 
Let us give global descriptions of anti--automorphisms in $N=1$ :  
\{(\ref{eq:4-anti-automorphisms}), Lemma 1\} and $N=2$ :   
\{(\ref{eq:9-a-anti-automorphisms}), Lemma 6\} for the readers who are 
not familiar with this field.
\vspace{3mm} 
\par \noindent
 (a) {\bf $N = 1$} : \quad 
   $\flat : G_{1}(\fukuso^2) \longrightarrow  G_{1}(\fukuso^2)$
\par \noindent 
Since 
\[
  P(z)=\frac{1}{1+\zetta{z}^2}
     \left(
         \begin{array}{cc}
             1& {\bar z} \\
             z& \zetta{z}^2 
         \end{array}
     \right) 
\]
we have 
\begin{eqnarray}
  &&P({\bar z})=\frac{1}{1+\zetta{z}^2}
     \left(
         \begin{array}{cc}
             1& z \\
             {\bar z}& \zetta{z}^2 
         \end{array}
     \right)= \overline{P(z)}, \quad  
  P(-{\bar z})=\frac{1}{1+\zetta{z}^2}
     \left(
         \begin{array}{cc}
             1& -z \\
             -{\bar z}& \zetta{z}^2 
         \end{array}
     \right), \nonumber \\ 
  &&P(1/{\bar z})=\frac{1}{1+\zetta{z}^2}
     \left(
         \begin{array}{cc}
             \zetta{z}^2& {\bar z} \\
             z& 1
         \end{array}
     \right), \quad  
  P(-1/{\bar z})=\frac{1}{1+\zetta{z}^2}
     \left(
         \begin{array}{cc}
             \zetta{z}^2& -{\bar z} \\
             -z& 1
         \end{array}
     \right), \nonumber 
\end{eqnarray}
so that it is easy to see 
\begin{eqnarray}
 \label{eq:global-4-anti-automorphisms}
 &&\mbox{(1)}\ \  \flat : P \longrightarrow {\bar P}\qquad  \qquad
  \mbox{(2)}\ \  \flat : P \longrightarrow \sigma_{3}{\bar P}\sigma_{3}
     \nonumber \\
 &&\mbox{(3)}\ \  \flat : P \longrightarrow \sigma_{1}{\bar P}\sigma_{1}
     \qquad  
  \mbox{(4)}\ \  \flat : P \longrightarrow 
          \sigma_{1}\sigma_{3}{\bar P}\sigma_{3}\sigma_{1} 
\end{eqnarray}
with $\{\sigma_{3}, \sigma_{1}\}$ in (\ref{eq:sigma31}). At this moment 
we know that anti--automorphisms are independent of choices of local 
coordinates.
\vspace{5mm}\par \noindent
 (a) {\bf $N = 2$} : \quad 
   $\flat : G_{1}(\fukuso^3) \longrightarrow  G_{1}(\fukuso^3)$
\par \noindent 
Since 
\begin{eqnarray}
  P(z_1,z_2)&=&\frac{1}{1+\zetta{z_1}^2+\zetta{z_2}^2}
     \left(
         \begin{array}{ccc}
             1& {\bar z_1}& {\bar z_2}  \\
             z_1& \zetta{z_1}^2& z_1{\bar z_2} \\
             z_2& z_2{\bar z_1}& \zetta{z_2}^2  
         \end{array}
     \right),  \nonumber \\
  P({\bar z_1},{\bar z_2})&=&\frac{1}{1+\zetta{z_1}^2+\zetta{z_2}^2}
     \left(
         \begin{array}{ccc}
             1& z_1& z_2  \\
             {\bar z_1}& \zetta{z_1}^2& {\bar z_1}z_2 \\
             {\bar z_2}& {\bar z_2}z_1& \zetta{z_2}^2  
         \end{array}
     \right)=\overline{ P(z_1,z_2)} \nonumber 
\end{eqnarray}
we have after some calculations 
\begin{eqnarray}
 \label{eq:global-9-anti-automorphisms}
 &&\mbox{(a--1)}\ \ \flat : P \longrightarrow {\bar P}\quad \qquad \ 
   \mbox{(a--2)}\ \ \flat : P \longrightarrow A{\bar P}A \quad \qquad \ 
   \mbox{(a--3)}\ \ \flat : P \longrightarrow A^2{\bar P}A^2  
   \nonumber \\
 &&\mbox{(b--1)}\ \ \flat : P \longrightarrow B{\bar P}B\quad \ \ 
   \mbox{(b--2)}\ \ \flat : P \longrightarrow BA{\bar P}AB \quad \ \ 
   \mbox{(b--3)}\ \ \flat : P \longrightarrow BA^2{\bar P}A^2B  
   \nonumber \\
 &&\mbox{(c--1)}\ \ \flat : P \longrightarrow B^2{\bar P}B^2\quad 
   \mbox{(c--2)}\ \ \flat : P \longrightarrow B^2A{\bar P}AB^2 \quad 
   \mbox{(c--3)}\ \ \flat : P \longrightarrow B^2A^2{\bar P}A^2B^2 
   \nonumber \\ 
 &&{} 
\end{eqnarray}   
with $\{A, B\}$ in (\ref{eq:AB}). We leave the proof to the readers. 
At this moment anti--automorphisms are independent of choices of local 
coordinates.

\vspace{3mm} \par \noindent 
The generalization of the above result to arbitrary $N$ is very easy,   
\cite{KF6}.
\vspace{1cm}
\par \noindent
\begin{Large}
  \textbf{Appendix B \ \ Walsh--Hadamard Transformation}
\end{Large}
\vspace{3mm} \par \noindent 
Let us make a brief comment on the Walsh--Hadamard transformation. 
See \cite{AH} and \cite{KF5} about it.

The Walsh--Hadamard transformation $W$ defined by 
\begin{equation}
  W=\frac{1}{\sqrt{2}}
     \left(
         \begin{array}{cc}
             1& 1 \\
             1& -1
         \end{array}
     \right) \quad \in \quad U(2)
\end{equation}
plays a very important role in quantum computation, see \cite{KF5}. This 
transformation is characterized as 
\begin{equation}
     \left(
         \begin{array}{cc}
              & 1 \\
             1& 
         \end{array}
     \right) = 
    W
     \left(
         \begin{array}{cc}
             1&  \\
              & -1
         \end{array}
     \right) 
    W^{-1}\ .
\end{equation}

\vspace{5mm}
Let us take an analogy. If we define $\widetilde{W}$ as 
\begin{equation}
   \widetilde{W}=\frac{1}{\sqrt{3}}
    \left(
         \begin{array}{ccc}
             1& 1&1  \\
             1& \omega^2& \omega \\
             1& \omega& \omega^2 \\
         \end{array}
     \right) \quad \in \quad U(3)
\end{equation} 
then we have 
\begin{equation}
     \left(
         \begin{array}{ccc}
             0 & & 1 \\
             1& 0&  \\
              & 1& 0 
         \end{array}
     \right) = 
    \widetilde{W}
     \left(
         \begin{array}{ccc}
             1&  &   \\
              & \omega&  \\
              & & \omega^2 
         \end{array}
     \right) 
   \widetilde{W}^{-1}\ .
\end{equation}

\par \noindent
Therefore it may be possible for us to call $\widetilde{W}$ generalized 
Walsh--Hadamard transformation.



\begin{thebibliography}{99}
%
\bibitem{LPS}H-K. Lo, S. Popescu and T. Spiller (eds) : 
\newblock Introduction to Quantum Computation and Information, 
\newblock 1998, World Scientific. 

%
\bibitem{JB}J. S. Bell : 
\newblock On the Einstein-Podolsky-Rosen paradox, 
\newblock Physics, 1(1964), 195.
%
\bibitem{APe}A. Peres : 
\newblock Quantum Theory : Concepts and Methods, 
\newblock 1995, Kluwer Academic Publishers. 
%
\bibitem{AH} A. Hosoya : 
\newblock Lectures on Quantum Computation (in Japanese), 
\newblock 1999, Science Company (in Japan).
%
\bibitem{MW}L. Mandel and E. Wolf : 
\newblock Optical Coherence and Quantum Optics,
\newblock 1995, Cambridge University Press. 
%
\bibitem{AP}A. Perelomov : 
\newblock Generalized Coherent States and Their Applications,
\newblock Springer--Verlag, 1986.
%
\bibitem{DF}D. I. Fivel : 
\newblock How a Quantum Theory Based on Generalized Coherent States 
Resolves the EPR and Measurement Problems, 
\newblock quant-ph/0104123.
%
\bibitem{FKSF1}K. Funahashi, T. Kashiwa, S. Sakoda and K. Fujii : 
\newblock Coherent states, path integral, and semiclassical approximation,
\newblock  J. Math. Phys., 36(1995), 3232.
%
\bibitem{FKSF2}K. Funahashi, T. Kashiwa, S. Sakoda and K. Fujii : 
\newblock Exactness in the Wentzel-Kramers-Brillouin approximation for 
some homogeneous spaces,
\newblock J. Math. Phys., 36(1995), 4590.
%
\bibitem{FKS}K. Fujii, T. Kashiwa, S. Sakoda :
\newblock Coherent states over Grassmann manifolds and the WKB exactness
in path integral,
\newblock J. Math. Phys., 37(1996), 567.
%
\bibitem{MN}M. Nakahara : 
\newblock Geometry, Topology and Physics,
\newblock IOP Publishing Ltd, 1990.
%
\bibitem{KF5}K. Fujii : 
\newblock Introduction to Grassmann Manifold and Quantum Computation, 
\newblock quant-ph/0103011.
%
\bibitem{KF6}K. Fujii : 
\newblock Geometry, Coherence and Entanglement (a tentative title),  
\newblock in progress. 
%
%
%
\bibitem{ZR}P. Zanardi and M. Rasetti : 
\newblock Holonomic Quantum Computation,
\newblock Phys. Lett. A264(1999), 94,
\newblock quant-ph/9904011.
%
\bibitem{PZR}J. Pachos, P. Zanardi and M. Rasetti : 
\newblock Non-Abelian Berry connections for quantum computation,
\newblock to appear in Phys. Rev. A, 
\newblock quant-ph/9907103.
%
\bibitem{PC}J. Pachos and S. Chountasis : 
\newblock Optical Holonomic Quantum Computer,
\newblock quant-ph/9912093.
%
\bibitem{PZ}J. Pachos and P. Zanardi : 
\newblock Quantum Holonomies for Quantum Computing,
\newblock quant-ph/0007110. 
%
%
\bibitem{KF1} K. Fujii : 
\newblock Note on Coherent States and Adiabatic Connections, Curvatures,
\newblock J. Math. Phys.,  
\newblock 41(2000), 4406.
%
\bibitem{KF2} K. Fujii : 
\newblock Mathematical Foundations of Holonomic Quantum Computer,
\newblock to appear in Rept. Math. Phys, 
\newblock quant-ph/0004102.
%
\bibitem{KF3} K. Fujii : 
\newblock More on Optical Holonomic Quantum Computer,
\newblock quant-ph/0005129.
%
\bibitem{KF4} K. Fujii : 
\newblock Mathematical Foundations of Holonomic Quantum Computer II,
\newblock quant-ph/0101102.
%
%
\end{thebibliography}
\end{document}